\documentclass[global, final,twocolumn]{svjour}
\usepackage[logonly]{trace}
\usepackage{graphicx}
\usepackage{amsmath}
\usepackage{amsfonts}
\usepackage{amssymb}
\usepackage{lineno}

\title{Cosmic vacuum energy decay and creation of cosmic matter}
\author{Hans-J\"org Fahr $\cdot$ Michael Heyl}  
\offprints{H.-J. Fahr \\
Argelander Institute for Astronomy \\
University of Bonn \\
53121 Bonn, Germany \\
email: hfahr@astro.uni-bonn.de \\ \\       
M. Heyl \\
German Space Agency - Department of Navigation \\
German Aerospace Center \\
53227 Bonn, Germany}

\date{Received: date / Revised version: date}

\begin{document}
\authorrunning{H.J. Fahr \& M. Heyl}
\titlerunning{Cosmic vacuum energy decay and creation of cosmic matter}
\maketitle


\begin{abstract}
In the more recent literature on cosmological evolutions of
the universe the cosmic vacuum energy has become a non-renouncable ingredient.
The cosmological constant $\Lambda$, first invented by Einstein, but later
also rejected by him, presently experiences an astonishing revival.
Interestingly enough it acts, like a constant vacuum energy density would also
do. Namely, it has an accelerating action on cosmic dynamics without which, as
it appears, presently obtained cosmological data cannot be conciliated with
theory. As we are going to show in this review, however, the concept of a
constant vacuum energy density is unsatisfactory for very basic reasons, since
it would claim for a physical reality that acts upon spacetime and matter
dynamics without itself being acted upon by spacetime or matter.


\end{abstract}
\section{Introduction}
\label{Intro}

If vacuum energy acts upon cosmic spacetime, then, doing so, it should also be
acted upon in some respect. Most probably when driving the expansion of the
universe vacuum energy should decay, but the question arises itself : How does
it decay? And does it decay into matter or something else? The question of how
matter creation and vacuum energy decay should be connected with each other has
not at all been answered satisfactorily up to the present. There exists,
however, a long series of papers, in which the connection between vacuum
energy decay and matter creation is touched and is brought to at least
tentative formulations. In the following we shall review a few of the most
prominent of these contextual formulations.

As it perhaps appears the most promising way to follow and push forward such
ideas is to continue the line of argumentations given by Hawking (1975) who in
form of his well known "Hawking radiation" has described quantummechanical
mechanisms to convert vacuum energy into radiations of matter. First of all,
in any case there are some eminent differences between mass energy density and
vacuum energy density which from the very beginning of any further studies
should not be overlooked: These conceptual differences are centered around the
fact that mass energy density is connected with massive particles present in a
unit volume of space, whereas vacuum energy density is a virtue of the volume
itself (see Overduin and Fahr, 2001, Fahr, 2004). The presence of
"something" always is connected with a specific place in configuration space
where this something in form of a single particle is present, i.e.
"topified". The vacuum, i.e. the absence of something, to the contrast is of
course not connected with a specific place. This makes a tremendous
difference. Redistributing the same number of particles in a larger volume
necessarily means that mass density of these particles decreases as a
reciprocal of the volume. Enlarging the volume of a vacuum, in contrast, does
not have an evident reaction in the vacuum energy density. If vacuum energy
for instance is taken to be constant, as done in many cases of the present
literature, then a vacuum of twice the volume simply represents twice the
energy. Nevertheless there exist many indications that vacuum energy density
should not be constant, but should decay with the expansion of the universe,
its energy equivalent most probably reappearing in form of a gain of mass density of
real matter.

Indications for that context were given for instance in a recent paper by Fahr
and Heyl (2006a/b) who have shown that an $S^{-2}$-scaling (with $S$ the extension of the universe) of both cosmic
vacuum energy density and mass energy density can guarantee a universe with
constant total energy. Furthermore, this behaviour can also conciliate
the completely disjunctive claims for the magnitude of the vacuum energy
density which are in the literature at present; on one hand very high
vacuum energy densities have been calculated by field theoreticians, while on
the other hand claims for very low vacuum energy densities originate from more
recent results of observational cosmology. A mass generation rate of half a Planck mass per Planck time for the expanding universe for instance nicely
explains the present material universe as coming from nothing but pure vacuum.
However, for that to be true it is necessary that both the cosmic mass density
and the cosmic vacuum energy density do scale with $S^{-2}$.

The question thereby remains, why cosmic spacetime expansion should trigger
the vacuum energy decay, and how the conversion of vacuum fluctuations into
real massive particles may occur. This is an outstanding question up to now,
not answered at all in a satisfactory manner in the scientific literature. At
least, however, strong reasons exist to believe that cosmic vacuum energy, if
it has an accelerating action on cosmic spacetime, has to decay. A constant
vacuum energy doing an action on space by accelerating its expansion, without
itself being acted upon, does not seem to be a concept conciliant with basic
physical principles. In this respect all conservative Lambda-cosmologies
taking the vacuum energy density as a constant (see e.g. models used in Perlmutter et al.,
1999, or Bennett et al., 2003) in our view are nonconvincing solutions. This
has meanwhile been realized by many authors like e.g. Kolb (1989), Fischer
(1993), Massa (1994), Wetterich (1995), Overduin and Cooperstock (1998), Fahr
(2004,2006) or Dutta-Choudhury and Sil (2006) where variable-Lambda
cosmologies have been discussed in detail.

The connection of vacuum energy and matter creation has already been discussed
in early papers by Hoyle (1948), Hoyle and Narlikar (1966 a/b) and later by
Hoyle (1990,1992) and Hoyle, Burbidge and Narlikar (1993). The requirement
that general relativistic field equations should be conformally invariant with
respect to any scale recalibrations leads these authors to the
introduction of a general relativistic action potential which describes mass
generation connected with geodetic motions of particles. To describe this form
of mass generation a so-called $C$-field (creation-field) can be introduced
which turns out to be connected with geodetic mass motions themselves. It can
then be shown (Hoyle, Burbidge, Narlikar, 1993) that this $C$- field when
introduced into the general relativistic field equations leads to terms
equivalent to those resulting from vacuum energy (see section 2 below). A similar connection between
vacuum energy density and mass density was also found by Massa (1994) who
shows that the cosmological term Lambda should be coupled to matter density,
concretely to be proportional to mass density, meaning that when the latter is
decreasing the former should also decrease.

We start our review with a look
into the cosmological literature of the past where cosmic mass generation
mechanisms had been formulated and can show there that various, theoretically
described forms of mass generation in the universe lead to terms in Einstein`s
field equations of general relativity which in many respects are analogous to,
or can even be replaced by terms arising from vacuum energy. We then analyse
other cosmological literature where it has been demonstrated that
gravitational cosmic binding energy acts as negative cosmic mass energy and,
as a surprise, again this leads to terms similar to those written for cosmic
vacuum energy. As it turns out from these studies the cosmic actions of vacuum
energy, gravitational binding energy and mass creation are evidently closely
related to each other. Based on these results we suggest that the action of
vacuum energy on cosmic spacetime necessarily and simultaneously leads to a
decay of vacuum energy density, a decrease of cosmic binding energy, and a
creation of mass in the expanding universe. We demonstrate that only under
these auspices an expanding universe with constant total energy, the so-called
economic universe, is possible. In such a universe both cosmic mass density
and cosmic vacuum energy density are decreasing according to $(1/S^{2})$ ,
with $S$ being the characteristic scale of the universe. Just under these
conditions the origin of the present universe from an initial, complete vacuum
reveals possible. The incredibly huge vacuum energy density which quantumfield
theoreticians do derive meanwhile has decayed to the small value which is
conciliant with the behaviour of the present universe, but reappears in the
energy density of created cosmic matter.

Concluding these considerations one can say that up to now there is surely a lack of a
rigorous formulation for the transition of vacuum fluctuations into real
masses, nevertheless there is at least the idea first discussed by Hawking
(1975) that treating the quantummechanics of particles and antiparticles in
the neighborhood of blackholes reveals the appearance of real particles, i.e.
the decay of a pure vacuum into a mass-loaded vacuum around these blackholes.
The so-called Hawking radiation in this respect is nothing else but a
materialisation of vacuum energy occuring in strong gravitational fields.
Perhaps in this respect the expanding universe also represents a form of a strongly
time-dependent gravitational field which induces matter creation through the
embedded time-dependent quantummechanical wavefunctions of particles.

\section{Basics of Hoyle`s creation theory}
\label{sec:1}

Hoyle (1948) was amongst the first to consider creation of mass in an
expanding universe. One of his motivations has been to describe a universe
which fulfills the cosmological principle in its strongest form, i.e. a
universe which not only looks alike from each spacepoint at one selected
cosmic time $t$, but looks alike also for all cosmic times, so that there
exist neither prefered cosmic places nor cosmic times what concerns the
observation of cosmic properties. For that purpose it was clear that matter
creation is needed in an expanding universe, since otherwise a systematic
decrease of mass density in the universe is the unavoidable consequence.

Following the ideo-aesthetical view of Weyl (1920) a field theory like GRT
with a metric of the Poincar$\acute{\mathrm{e}}$-Weyl group can be scale-invariant and, - connected with the minimum action
principle which should be always fulfilled -, then requires that mass is
created at geodetic motions of comoving cosmic masses. This leads Hoyle (1948)
to the introduction of a new 2-rank tensor field $C_\mathrm{\mu\nu}$ which is derived
as the covariant derivative of a time-like geodesic 4-vector $C_\mathrm{\mu
}=3c/A\{1,0,0,0\}$ defined at each cosmic place with some constant scale
factor $A$. Thus Hoyle obtains this tensor field in the following form:%

\begin{equation}
C_\mathrm{\mu\nu}=\frac{\partial C_\mathrm{\mu}}{\partial x^{\nu}}-\Gamma_\mathrm{\mu\nu}^{\alpha
}C_\mathrm{\alpha} \label{1}
\end{equation}

When the above Christoffel symbols $\Gamma_\mathrm{\mu\nu}^{\alpha}$ are evaluated on the
basis of the Robertson-Walker metric, then it is found that only tensor
elements with $\mu,\nu \ne 0$ are non-vanishing and are given by:%

\begin{equation}
C_\mathrm{\mu\nu}=-3S\dot{S}\frac{\delta_\mathrm{\mu\nu}}{cA} \label{2}
\end{equation}

for $\mu,\nu=1,2,3$. Here $\delta_\mathrm{\mu\nu}$ means the Konecker symbol.

Now Hoyle (1948) with the use of the $C_\mathrm{\mu\nu}$ field, with some inherent
arbitrariness, completes the GRT\ field equations\ with a term, containing
purely metrical quantities, on their left hand sides by writing:%

\begin{equation}
G_\mathrm{\mu\nu}-\frac{1}{2}g_\mathrm{\mu\nu}G+C_\mathrm{\mu\nu}=\frac{8\pi\gamma}{c^{4}}T_\mathrm{\mu
\nu} \label{3}
\end{equation}

From that system of equations he obtains for the nontrivial cosmological relations:%

\begin{equation}
2S\ddot{S}+\dot{S}^{2}-3cS\dot{S}\frac{1}{A}=0 \label{4}
\end{equation}

and:%

\begin{equation}
3\dot{S}^{2}=8\pi\gamma\rho S^{2} \label{5}
\end{equation}

From these two differential equations one then obtains the following solutions
(given in normalized quantities, i.e. with $S=1$ at $t=0$!$)$:%

\begin{equation}
S=\exp[ct/A] \label{6}
\end{equation}

and:%

\begin{equation}
\rho=\rho_\mathrm{H}=\frac{3c^{4}}{8\pi\gamma A^{2}} \label{7}
\end{equation}

which means a de Sitter-type inflationary expansion of the universe and a
constant density, connected with a mass generation rate (source strength) given by:%

\begin{equation}
\dot{\sigma}_\mathrm{\rho}=3\frac{\dot{S}}{S}\rho_\mathrm{H}=3\frac{c}{A}\rho_\mathrm{H} \label{8}
\end{equation}

For the matter-less (i.e. $\rho=0$) de Sitter cosmology with a cosmological
constant $\Lambda$ (see Einstein and de Sitter, 1932) also an
inflationary expansion is found which leads to an expansion law given by:%

\begin{equation}
S=\exp[\sqrt{\frac{\Lambda c^{2}}{3}}t] \label{9}
\end{equation}

suggesting that mass generation, according to the style Hoyle describes it,
could be taken as analogous to a cosmological constant of the form:
$\Lambda_\mathrm{H}=3/A^{2}$. One should, however, clearly see the difference, since
Hoyle`s mass creation tensor field $C_\mathrm{\mu\nu}$ induces inflation, though the
tensor element $C_\mathrm{00}$ vanishes, i.e. $C_\mathrm{00}=0!$, whereas de Sitter`s vacuum
cosmology just induces inflation by that corresponding vacuum energy-momentum
tensor element being different from zero, i.e. $T_\mathrm{00}=\Lambda\ne 0!$.
Disregarding this difference one could feel inclined to see in this approach
some indication of an equivalence of mass creation and vacuum energy action
given by the relation:

\begin{equation}
\dot{\sigma}_\mathrm{\rho}=3\frac{c}{A}\rho_\mathrm{H}=\frac{9c^{5}}{8\pi\gamma A^{3}}=\frac
{c^{5}\Lambda_\mathrm{H}^{3/2}\sqrt{3}}{8\pi\gamma} \label{10}
\end{equation}

It is also interesting to see the influence on the light horizon coming up in
Hoyle`s universe. Connected with the above relation for the scale factor, one
easily obtains for the distance from an arbitrary spacepoint a photon emitted
at time $t_\mathrm{1}$ can reach at maximum in time ($t\rightarrow\infty$)\ the
following result:%

\begin{equation}
r_\mathrm{1,\infty}=A\cdot\lbrack e^{-ct_\mathrm{1}/A}- e^{-ct_\mathrm{\infty}/A}]=A\cdot
e^{-ct_\mathrm{1}/A} \label{11}
\end{equation}

This also means that a photon emitted from beyond this distance $r_\mathrm{1,\infty}$ at
time $t_\mathrm{1}$ can never reach us (Hoyle 1948).

\section{Gravitational binding energies as hidden masses}
\subsection{The zero-energy universe (economical universe)}

The idea that the dynamic action of a cosmic vacuum is coupled with the cosmic
material environment appears highly suggestive. Imagine for instance a small
volume of vacuum surrounded by a barostatically structured material
environment embedded in a large scale gravitational field. It is evident then
that the way how the vacuum bubble couples to the gravitational field, if not
due to a direct genuine force upon the vacuum, is simply due to the buoyancy
force exerted to the vacuum bubble by the pressure distribution of the ambient
matter. This pressure of course depends on the matter density.

The universe, however, is not in a barostatic state, but rather in the state
of a dynamic equilibrium with conserved total cosmic energy. Thus one may find
the relation between vacuum energy decay and material mass density
evolution along a little bit different way: We assume that, for
respecting a kind of cosmic economy principle, the total energy of the
universe should be conserved during cosmic evolution and should vanish.
This concept was already earlier formulated at many places in the
literature (see e.g. Tryon, 1973, Brout et al., 1978, Vilenkin, 1982, Rosen,
1994, Cooperstock and Israelit, 1995, Overduin and Fahr, 2001). Following this
idea one can then investigate consequences this might have for the way how
vacuum energy and matter density should be related to eachother.

In what follows we present a calculation with a volume scale $r\leq
S$, where $S\simeq c/H$ is the scale of the universe
and $H$ is the well known Hubble constant, tacitly assuming that this
estimate can be extended to the whole universe. On the scale $S$ no
spacetime curvature, if at all present, may need be taken into account, and a
quasi-Euclidean metrics can be used. Thus the energy $E$ of all
masses in this sub-universe is given by:%

\begin{equation}
E=\int^{V^{3}}(\rho c^{2}+3p)\sqrt{-g_{3}}d^{3}V=\frac{4\pi}{3}S^{3}(\rho c^{2}+3p) \label{12}
\end{equation}

Here $\sqrt{-g_{3}}d^{3}V$ is the local
differential proper volume of 3-d space given through the determinant of the
3-d space-like part of the Robertson-Walker metric tensor which for vanishing
curvature, i.e. $k=0$, leads to the outer right result in the above
equation (for details see Fahr and Heyl, 2006b).

In the above expression the total mass density is given by:

\begin{equation}
\rho=\rho_\mathrm{b}+\rho_\mathrm{d}+\rho_\mathrm{vac} \label{13}
\end{equation}

and the total pressure is given by:

\begin{equation}
p=p_\mathrm{b}+p_\mathrm{d}+p_\mathrm{vac} \label{14}
\end{equation}

with the indices b, d, vac denoting quantities of baryonic
matter, dark matter and the vacuum, respectively.

In the present phase of the evolution of the universe baryonic and
dark matter can be considered as cold and pressure-less, i.e. $p_\mathrm{b}+p_\mathrm{d}%
=0$. Assuming furthermore a general dependence of $\rho_\mathrm{vac}\sim
S^{-n}$ (see Fahr and Heyl, 2006b) one then obtains: $p=p_\mathrm{vac}%
=-\frac{3-n}{3}\rho_\mathrm{vac}c^{2}$ and finds:

\begin{equation}
E=\frac{4\pi}{3}S^{3}c^{2}(\rho_\mathrm{b}+\rho_\mathrm{d}+(n-2)\rho_\mathrm{vac}) \label{15}
\end{equation}

In addition to the positive-valued energy $E$ one has to put
into the balance the negative-valued gravitational potential binding energy
$U$ to be found from the gravitational cosmic potential $\Phi$.

The latter, as we first assume here, can be calculated with the help
of the well known Poisson equation of the cosmic potential through

\begin{equation}
\Delta\Phi=\frac{1}{r^2}\frac{\partial}{\partial r}(r^2\frac{\partial}{\partial r}\Phi)=-4\pi G(\rho_\mathrm{b}+\rho
_\mathrm{d}+(n-2)\rho_\mathrm{vac}) \label{16}
\end{equation}

where $r$ is a reference point related radial coordinate. In
a homogeneous universe this equation is then solved by:

\begin{equation}
\Phi(r)=-\frac{2}{3}\pi G(\rho_\mathrm{b}+\rho
_\mathrm{d}+(n-2)\rho_\mathrm{vac})r^{2} \label{17}
\end{equation}

yielding the total amount of gravitational binding energy in the
following form:

\begin{equation}
U=\int_{0}^{S}4\pi r^{2}(\rho_\mathrm{b}+\rho_\mathrm{d}+(n-2)\rho_\mathrm{vac})\Phi(r)dr \label{2100}
\end{equation}

which leads to the expression:

\begin{equation}
U =-\frac{8\pi^{2}%
G}{15}(\rho_\mathrm{b}+\rho_\mathrm{d}+(n-2)\rho_\mathrm{vac})^{2}S^{5} \label{2101}
\end{equation}

It is perhaps very interesting, and also relaxing at the same time, to
recognize that the above result could be as well derived with the use of the
effective cosmic potential

\begin{equation}
\Phi_\mathrm{eff}=\frac{8\pi G}{3} (\rho_\mathrm{b}+\rho_\mathrm{d}+(n-2)\rho_\mathrm{vac})S^{2} \label{19}
\end{equation}

derived by Fahr and Heyl (2006b) in their Eq.(2) and yielding as the
only difference with respect to the above Eq. (\ref{2101}) a factor $\frac{20}{3}$.

Now the requirement that the total energy of the universe $L=E+U$ vanishes - with $E$ and $U$ given by Eqs. (\ref{15}) and (\ref{2101}) - thus leads to the explicit requirement that either:

\begin{equation}
(\rho_\mathrm{b}+\rho_\mathrm{d}+(n-2)\rho_\mathrm{vac})=0 \label{21}
\end{equation}

or that:

\begin{equation}
\frac{3c^{2}}{2\pi GS^{2}}=(\rho_\mathrm{b}+\rho_\mathrm{d}+(n-2)\rho_\mathrm{vac}) \label{22}
\end{equation}

The first of these two relations can only be fulfilled for all values
of $S$, if all quantities have the same $S$-dependence and
can compensate to zero, which would only be possible for $n\leq2$.
The second relation can, however, always be fulfilled, if the common
scale-dependence of $\rho_\mathrm{b}, \rho_\mathrm{d}, \rho_\mathrm{vac}$ is given by a
$\left(1/S^{2}\right)$-dependence, necessarily meaning that
$n=2$.

It could appear that the required $\left(1/S^{2}\right)$-dependence of the mass densities $\rho_\mathrm{b}$ and $\rho_\mathrm{d}$
does not allow to fulfill the normal conservation law for masses, since the
latter should lead to a $\left(1/S^{3}\right)$-dependence. This
is, however, not true in our case here where matter creation by vacuum decay
has to be taken into account, because under these conditions the conservation
law writes%

\begin{equation}
\dot{S}\frac{d}{dS}\rho_\mathrm{mat}+\frac{1}{S^{2}}\frac{d}{dS}(S^{2}\rho_\mathrm{mat}\dot{S})=\dot{S}%
\frac{d}{dS}\rho_\mathrm{vac} \label{2102}
\end{equation}

which under the conditions given above, i.e.  $\frac{d}{dS}\rho
_\mathrm{mat}=\frac{d}{dS}\rho_\mathrm{vac}$, fulfilled in the economic universe,
then simply leads to the result $S^{2}\rho_\mathrm{mat}\dot{S}=const$, and
thus with Equ.(70) simply yields $\rho_\mathrm{mat}=const/(cS^{2})$.

The relation between $\rho_\mathrm{b}, \rho_\mathrm{d}, \rho_\mathrm{vac}$ must be realized in such a way as to fulfill the $\left(1/S^{2}\right)$-dependence for all of them. If the number of
individual baryonic and dark particles is conserved and their masses are
invariable, then the particle densities should fall off with $\left(
1/S^{3}\right)$. This means that consequently the needed $\left(
1/S^{2}\right)$-dependence is only realized, if either the baryonic
and dark particle masses $m_\mathrm{b}$ and $m_\mathrm{d}$ themselves both
vary proportional to $R$, or if their number changes due to some
well-tuned particle creation rate (see e.g. Vishwakarma, 2003, Unnikrishnan et
al., 2002, Fahr and Heyl, 2006b).

The required $1/S^{2}-$scaling of the densities appears in
fact a bit embarassing and intriguing at first glance, since meaning that the
''economic universe'' by this point seems to be in conflict with most recent
observational results telling that the universe is at present accelerating its
expansion and not coasting with $\dot{S}=c$. But to tell the truth,
the cosmological parameters derived from the WMAP (Bennet et al., 2003) and
the distant SN (Perlmutter et al., 1998) data, are not observational facts by
themselves. They are, one should clearly see that, highly ''ideology-loaded''
best fit data representations, applicable only on the basis of the standard
cosmological Friedman model, enlarged by the admission of terms for dark
matter and dark energy. But it must be seen that these cosmological model
parameters are found assuming that $\rho_\mathrm{b}$, $\rho_\mathrm{d}$
scale with $S^{-3}$, and that $\rho_\mathrm{vac}$ is constant which
is completely different in the economical universe presented here. In the near
future we shall have to show how WMAP data can be understood on the basis of
this new economic cosmological model.

The fact, however, that a ratio of $\rho_\mathrm{vac}/\rho_\mathrm{mat}\simeq
7/3$ as pointed out by conventional WMAP interpretations can
nevertheless give a good hint, because in our theory presented here this ratio
should not only be valid for the present epoch of cosmic evolution - going up
to larger and larger values in the future (to smaller and smaller values in
the past) of the universe, but it should be a constant for the whole cosmic
evolution. So it would not at all be an ''anthropic miracle'' that we find
vacuum energy density and matter density in about the same magnitudes just at
our living times.

The mass increase according to our logics here should be ascribed to the
coupling of these particles to the vacuum density requiring that the following
mass relation has to be valid:%

\begin{equation}
m_\mathrm{b,d}=m_\mathrm{b,d}^{0}\sqrt{\frac{\rho_\mathrm{v0}}{\rho_\mathrm{vac}}} \label{23}
\end{equation}

This tight relation between the properties of real matter and vacuum appears
very astonishing, in a first glance poorly inviting and hardly conceivable, but when looking back
into the history of the philosophy of matter and vacuum already developed at
ancient greek philosophers it may eventually and somehow appear more familiar
to us (see e.g. Fahr 2004). It may in addition attain an unexpectedly deep modern sense in view of
present day physics where real particles are seen to polarize the vacuum and
the energy density of the latter is due to the intensity of this polarisation:
the more matter in space the more polarized and energetic is the vacuum of this space.

\subsection{Binding energy acting as vacuum energy}

In an interesting, but fairly forgotten paper Fischer (1993) takes up the idea
that the source of gravity, i.e. of the metric, - the so-called
energy-momentum tensor\ $T_\mathrm{\mu\nu}$ -, should contain all relevant forms of
energy, thus including also gravitational binding energy represented in an
adequate general-relativistic form. The argument for that to be true is that
the covariant divergence of both the Einstein tensor $G_\mathrm{\mu\nu}$ and the
source tensor $T_\mathrm{\mu\nu}$ should vanish. The latter, however, can physically
only be true, if $T_\mathrm{\mu\nu}$ is a conserved quantity. Neither mass energies
nor pressure energies are conserved quantities by themselves, only the total
energy including gravitational binding energy is such a conserved quantity.
Consequently the source tensor $T_\mathrm{\mu\nu}$ should include a term describing
binding energy.

For instance, taking all the cosmic mass included in an Einstein-Straus vacuole
of radius $R_\mathrm{ES}$ around a star and condensing it (Sch\"ucking 1954) to the central star of mass
$M=(4\pi/3)R_\mathrm{ES}^{3}\rho_\mathrm{c}$ would lead to a field source which not only
gravitates by its mass $M$, but also by its pressure energy. If negative
stellar binding energy would not compensate for this positive pressure energy,
then the universe of point masses would metrically be very different from the
Friedmann-Robertson-Walker universe in which these point masses have simply be
considered as spread out to a homogeneous material substrate. Consequently the
description of spacetime metrics can only then be hoped for to satisfactorily
take into account the above mentioned phenomenon, if gravitational selfenergy,
or potential energy, is included as a term in the energy-momentum tensor.

Fischer (1993) proposes to introduce into $T_\mathrm{\mu\nu}$ the term for potential
energy in the following form:%

\begin{equation}
T_\mathrm{\mu\nu}^{p}=-C\frac{\rho_\mathrm{c}}{\Gamma}g_\mathrm{\mu\nu} \label{24}
\end{equation}

where $\rho_\mathrm{c}$ is the cosmic mass density, $\Gamma$ is the cosmic curvature
radius, and $C$ is a constant the magnitude of which is essentially kept open,
but can be fixed such that a static universe like the one described by
Einstein (1973) is guaranteed. Introducing this term, instead of the
cosmological $\Lambda$- term, into Einstein`s static field equations for the
curvature parameter $k=1$ he finds the following two remaining differential equations:%

\begin{equation}
-\frac{2}{\Gamma^{2}}=\kappa(\frac{\rho_\mathrm{c}^{2}c^{2}}{2}+C\frac{\rho_\mathrm{c}%
}{\Gamma}) \label{25}
\end{equation}

and:%

\begin{equation}
0=-\kappa(\frac{\rho_\mathrm{c}^{2}c^{2}}{2}-C\frac{\rho_\mathrm{c}}{\Gamma}) \label{26}
\end{equation}

where $\kappa=4\pi G/c^{2}$ , and from the above obtains the following relations:

\begin{equation}
C=\frac{\rho_\mathrm{c}}{2} \label{27}
\end{equation}

and: 

\begin{equation}
\Gamma=\sqrt{\frac{c^{2}}{4\pi\rho_\mathrm{c}G}} \label{28}
\end{equation}

The static universe aimed at by Einstein, in this view, is nothing else but a
universe in which potential energy just balances mass energy. In this respect
it also can be concluded that the cosmological term, included by Einstein and
identified with negative pressure, in case of a static universe is completely
replaced by the term introduced by Fischer (1993) for potential energy , i.e.
$\Lambda=-C\frac{\rho_\mathrm{c}}{\Gamma}$.

In case of the time-dependent Friedmann-Robertson Walker universe, the
potential energy term introduced by Fischer (1993) analogously attains the
form $T_\mathrm{\mu\nu}=-C\frac{\rho_\mathrm{c}}{S}g_\mathrm{\mu\nu}$ and for pressure-less matter
leads to the following two field equations:%

\begin{equation}
\frac{c^{2}k}{S^{2}}+\frac{\dot{S}^{2}}{S^{2}}+2\frac{\ddot{S}}{S}%
=\frac{\kappa C\rho_\mathrm{c}}{S} \label{29}
\end{equation}

and:

\begin{equation}
-3(\frac{c^{2}k}{S^{2}}+\frac{\dot{S}^{2}}{S^{2}})=-\kappa\rho_\mathrm{c}%
-\frac{\kappa C\rho_\mathrm{c}}{S} \label{30}
\end{equation}

Setting $k=1$ and calling $S_\mathrm{0}$ the solution for the static universe, i.e.
for the case $\dot{S}=\ddot{S}=0$, one obtains from the above system of
equations one differential equation describing the time-dependence of the
scale factor in the form:%

\begin{equation}
\frac{\ddot{S}}{S}=\frac{\kappa\rho_\mathrm{c}}{6}(\frac{S_\mathrm{0}}{S}-1) \label{31}
\end{equation}

This shows that for $S\leq S_\mathrm{0}$ the acceleration $\ddot{S}$ is positive like
in case of a positive vacuum energy acting, while for $S\geq S_\mathrm{0}$ it is
negative like in case of a negative vacuum energy acting, leading to a kind of
oscillatory behaviour of the scale factor. In any case for small values of $S$
the potential energy dominates over mass energy which leads to a positive
acceleration and removes the possibility of a collapse. By the way the
collapse of stellar matter towards a black hole would also be impeded by this term.

\section{An attempt to introduce an effective mass density}

We recall the fact that a star undergoes a mass deficit given by the mass
equivalent of the gravitational binding energy of the stellar mass of the
order of $\delta M\simeq(r_\mathrm{B}/r_\mathrm{st})\cdot M$ where $M$ is the mass of the
star, and $r_\mathrm{B}$ and $r_\mathrm{st}$ are the stellar Schwarzschild radius and the
stellar radius, respectively. This means that the mass of a star or of a
galaxy is less than the number of proton masses constituting this star or the
galaxy. What gravitationally acts to the outside world and inertially reacts
to accelerations is in fact the reduced mass $M^{\ast}=M-\delta M$. Hence
looking for the effective sources of the metrics of the universe one cannot
simply identify them with the proper energy-momentum density, but with its
''effective'' density obtained after reduction of its selfbinding energy
equivalent. What does this mean for the effective cosmic matter density
$\rho^{\ast}$?

In a linearized treatment we can try to formulate the effective density by:%

\begin{equation}
\rho^{\ast}=\rho_\mathrm{0}-\frac{G}{c^{2}}\rho_\mathrm{0}^{2}\int_\mathrm{0}^{r_\mathrm{1}}\frac{(4\pi
r^{2}dr)(4\pi r^{3}/3)}{r} \label{32}
\end{equation}

where $\rho_\mathrm{0}$ is the bare proper mass density (i.e. number of proton masses
per unit of volume in an inertial reference system) without gravitational
selfbinding taken into account. Solving this integral for $r_\mathrm{1}^{3}=1cm^{3}$,
i.e. for the unit volume $\sqrt[3]{r_\mathrm{1}}=1,$ one obtains:%

\begin{equation}
\rho^{\ast}=\rho_\mathrm{0}(1-\frac{16}{15}\frac{\pi^{2}G\rho_\mathrm{0}}{c^{2}}) \label{33}
\end{equation}

This approximation may be extended to the nonlinear selfbinding limit by the
expression (see Fahr and Zoennchen, 2006):%

\begin{equation}
\rho^{\ast}=\rho_\mathrm{0}\exp(-\frac{16}{15}\frac{\pi^{2}G\rho_\mathrm{0}}{c^{2}}%
)=\rho_\mathrm{0}\exp(-\alpha\rho_\mathrm{0}) \label{34}
\end{equation}

with $\alpha=16\pi^{2}G/15c^{2}$ , which for small values of $\rho_\mathrm{0}$ again
delivers the above given linear approximation, while for high values of
$\rho_\mathrm{0}$ the above formulation would practically describe the vanishing of
the gravitational source strength of matter, i.e. just close to so-called
Big-bang the effective gravity would dissolve itself due to disappearance of
its sources. As it, however, turns out when looking at the above result in
quantitative terms, only in the highest density phases of the cosmic evolution
with $\rho_\mathrm{0}\geq10^{28}g/cm^{3}$ this selfbinding effect of matter would
become relevant, e.g. for black holes, whereas for more diluted matter phases
in the course of the later cosmic expansion this cosmic selfbindng becomes
irrelevant. This above approach may look a little too simple-minded, since it
started out from linear mass deficit calculations. Therefore in the following
we shall look a little deeper into the general weakness of the concept of density in
curved spacetimes.

It is well known that in gravitational fields there of course always exist
local free-falling (LFF) systems which represent inertial reference systems.
These LFF - systems are, however, only valid approximations in the
infinitesimal neighbourhood of the origin of such a system. At finite
distances of the origin tidal gravitational forces are acting which cause
metric deviations from the Minkowskian spacetime. Under such circumstances it
is hard to imagine how spatially extended volumes, like the unit volume, can
become proper volumes which could help defining proper densities. The devil`s
circle is that spacevolumes on the one hand can only be defined on the basis
of a known spacetime metrics, while on the other hand the metrics in GRT is
defined through mass densities. We intend to approach this problem along the
following line of argumentations:

From the work of Einstein and Straus (1945) one learns that single stellar
masses $M$ can be embedded in the global FRW metric,
connected with a cosmic mass density of smeared out stellar masses, when ascribing
to them a Schwarzschild vacuole which merges into the outer metric at the
so-called Einstein-Straus radius $R_\mathrm{ES}$ (see Sch\"ucking 1954). Using this concept one may
introduce an effective mass density $\rho^{\ast}$which is given by:%

\begin{equation}
\rho^{\ast}=\frac{M}{V_\mathrm{ES}^{3}}=\frac{\frac{4\pi}{3}\rho_\mathrm{o}R_\mathrm{ES}^{3}%
}{V_\mathrm{ES}^{3}} \label{35}
\end{equation}

Here $\rho_\mathrm{0}$ is the so-called proper density, and $V_\mathrm{ES}^{3}$ is the
spacelike 3-d volume inclosed into the Einstein-Straus vacuole. This volume
within the ES vacuole is to be calculated on the basis of the inner
Schwarzschild metric as if the substrate of the cosmic mass density would fill
this vacuole (see Stephani, 1988):%

\begin{equation}
V_\mathrm{ES}^{3}=\int^{3}\hspace{0.15cm}\sqrt{-g_\mathrm{3}}d^{3}V \label{36}
\end{equation}

where $\rho_\mathrm{0}(t)$ denotes the homogeneous cosmic mass density which is
variable with cosmic time $t$ , and where $\sqrt{-g_\mathrm{3}}d^{3}V$ is
the local differential proper volume of configuration space given through the
determinant of the 3-d part of the inner Schwarzschild metric in the form (see Fahr and Heyl 2006)

\begin{equation}
\sqrt{-g_\mathrm{3}}=\sqrt{-g_\mathrm{rr}g_\mathrm{\vartheta\vartheta}g_\mathrm{\varphi\varphi}}%
=\sqrt{\exp(\lambda(r)} \label{37}
\end{equation}

with%

\begin{equation}
\exp(-\lambda(r))=1-\frac{8\pi G}{rc^{2}}\rho_\mathrm{0}\int_\mathrm{0}^{r}x^{2}%
dx=1-\frac{8\pi Gr^{2}}{3c^{2}}\rho_\mathrm{0} \label{38}
\end{equation}

This then leads to the following volume:%

\begin{equation}
V_\mathrm{ES}^{3}=4\pi\int_\mathrm{0}^{R_\mathrm{ES}}\frac{r^{2}dr}{\sqrt{1-\frac{8\pi Gr^{2}%
}{3c^{2}}\rho_\mathrm{0}}} \label{39}
\end{equation}

For a universe with vanishing curvature, i.e. for $k=0$, one finds:%

\begin{equation}
V_\mathrm{ES}^{3}=4\pi(\frac{3c^{2}}{8\pi G\rho_\mathrm{0}})^{3/2}\int_\mathrm{0}^{\xi_\mathrm{ES}}%
\frac{\xi^{2}d\xi}{\sqrt{1-\xi^{2}}} \label{40}
\end{equation}

where $\xi$ and $\xi_\mathrm{ES}$ have been introduced by:

\begin{equation}
\xi=\sqrt{\frac{8\pi G\rho_\mathrm{0}}{3c^{2}}}r \label{41}
\end{equation}

and: 

\begin{equation}
\xi_\mathrm{ES}=\sqrt{\frac{8\pi G\rho_\mathrm{0}}{3c^{2}}}R_\mathrm{ES} \label{42}
\end{equation}

Reminding that:

\begin{equation}
\int_\mathrm{0}^{\xi_\mathrm{ES}}\frac{\xi^{2}d\xi}{\sqrt{1-\xi^{2}}}  = \frac{1}{2}\arcsin\xi_\mathrm{ES}-\frac{\xi_\mathrm{ES}}{2}\sqrt{1-\xi_\mathrm{ES}^{2}} \label{43}
\end{equation}

then yields with Eqs. (\ref{35}) and (\ref{40}) the following result for the effective density:

\begin{equation}
\rho^{\ast}=\frac{\frac{4\pi}{3}\rho_\mathrm{o}R_\mathrm{ES}^{3}}%
{4\pi(\frac{3c^{2}}{8\pi G\rho_\mathrm{0}})^{3/2}[\frac{1}{2}\arcsin\xi_\mathrm{ES}%
-\frac{\xi_\mathrm{ES}}{2}\sqrt{1-\xi_\mathrm{ES}{}^{2}}]} \label{44}
\end{equation}

From the above one simply obtains:%

\begin{equation}
\rho^{\ast}=\frac{\rho_\mathrm{o}\ \xi_\mathrm{ES}^{3}(\frac{8\pi G\rho_\mathrm{0}}{3c^{2}}%
)^{-3/2}}{3(\frac{3c^{2}}{8\pi G\rho_\mathrm{0}})^{3/2}[\frac{1}{2}\arcsin\xi
_\mathrm{ES}-\frac{\xi_\mathrm{ES}}{2}\sqrt{1-\xi_\mathrm{ES}{}^{2}}]} \label{45}
\end{equation}

which finally results in:

\begin{equation}
\rho^{\ast}=\rho_\mathrm{0}\frac{\xi_\mathrm{ES}^{3}%
}{\frac{3}{2}[\arcsin\xi_\mathrm{ES}-\xi_\mathrm{ES}\sqrt{1-\xi_\mathrm{ES}{}^{2}}]} \label{46}
\end{equation}

Reminding now that for a vanishing curvature parameter, i.e. $k=0$, one finds
with the second Friedmann equation: $H=$ $\dot{S}/S=\sqrt{8\pi G\rho_\mathrm{0}/3}$
the following relation:

\begin{equation}
\xi_\mathrm{ES}=\sqrt{\frac{8\pi G\rho_\mathrm{0}}{3c^{2}}}R_\mathrm{ES}=\frac{\dot{S}}%
{cS}R_\mathrm{ES}=\frac{SH}{c}\frac{R_\mathrm{ES}}{S}=\frac{R_\mathrm{ES}}{S} \ll1 \label{47}
\end{equation}

where we have defined the radius of the universe as the radius of the Hubble Sphere,
i.e. $c=HS$. One then can see that the effective density $\rho^{\ast}$
does change as a function of the ratio of the Einstein-Straus radius $R_\mathrm{ES}$
and the radius $S$ of the universe. 

Simplifying the integral $\int_\mathrm{0}^{\xi_\mathrm{ES}}\frac{\xi^{2}d\xi}{\sqrt
{1-\xi^{2}}}$ in Eq. (\ref{43}) for small arguments $\xi$ by a linear expansion of the
nominator leads to %

\begin{equation}
\int_\mathrm{0}^{\xi_\mathrm{ES}}\frac{\xi^{2}d\xi}{\sqrt{1-\xi^{2}}}\simeq\int_\mathrm{0}%
^{\xi_\mathrm{ES}}(1+\frac{1}{2}\xi^{2})\xi^{2}d\xi=\frac{1}{3}\xi_\mathrm{ES}^{3}+\frac
{1}{10}\xi_\mathrm{ES}^{5} \label{48}
\end{equation}

and thus gives the result:%

\begin{equation}
\rho^{\ast}=\rho_\mathrm{0}\frac{\xi_\mathrm{ES}^{3}}{3[\frac{1}{3}\xi_\mathrm{ES}^{3}+\frac{1}%
{10}\xi_\mathrm{ES}^{5}]}=\rho_\mathrm{0}\frac{1}{1+\frac{3}{10}\ \xi_\mathrm{ES}^{2}} \label{49}
\end{equation}

As one can see, the effective density $\rho^{\ast}$should approach the proper
density $\rho_\mathrm{0}$ for $\xi_\mathrm{ES}\rightarrow0$. On the other hand the larger
$\xi_\mathrm{ES}$ grows, the more reduced is the former with respect to the latter.
Reminding in addition that:

\begin{equation}
\xi_\mathrm{ES}=\sqrt{\frac{8\pi G\rho_\mathrm{0}}{3c^{2}}}R_\mathrm{ES}=\Psi\rho_\mathrm{0}^{1/6} \label{50}
\end{equation}

where $\Psi$ calculates to:%

\begin{equation}
\Psi=\sqrt{\frac{8\pi G}{3c^{2}}}\sqrt[3]{\frac{3M}{4\pi}} \label{51}
\end{equation}

one thus, for $\xi_\mathrm{ES}\ll1$ and for constant stellar masses $M$ embedded in an
expanding universe, finds the result:%

\begin{equation}
\rho^{\ast}=\rho_\mathrm{0}\frac{1}{1+\frac{3}{10}\ \xi_\mathrm{ES}^{2}}\simeq\rho_\mathrm{0}%
\cdot(1-\frac{3}{10}\Psi^{2}\rho_\mathrm{0}^{1/3}) \label{52}
\end{equation}

showing that the effective mass density is reduced the more, the higher is the
proper density. It is perhaps worth noting that with the above result we
more or less come back to the more easy-minded derivation of the effect of
binding energy presented in the beginning of this section.

\section{ Indications for stellar mass increase}

In a recent paper Fahr and Siewert (2006,2007) have studied frequency shifts of
photons freely propagating through the local spacetime of our solar system. It
turns out that the well known radiowave frequency shift observed at signals
reflected to the earth from the PIONEER-10/11 spacecraft (i.e. the so-called
PIONEER anomaly discussed in Anderson et al. 1998) can nicely be explained as well by its magnitude and its sign (blueshift) within a local Schwarzschild metric of the central solar mass $M$ which acts as increasing with cosmic time according to the following rate: $\dot{M}/M=H_\mathrm{0}$. Though, the required mass increase seems hard to advocate for, there is nevertheless a physical basis available.

One can easily study then what this mass increase means in terms of the mass
of the unperturbed cosmic matter density which is rejected from the local
Schwarzschild vacuole (see Einstein \& Straus 1945 or Schuecking 1954).
Starting from the Schuecking relation the central mass $M$ is associated with
a Schwarzschild vacuole of radius $R_\mathrm{ES}$ by the following relation:%

\begin{equation}
M=\frac{4\pi}{3}\rho_\mathrm{0}R_\mathrm{ES}^{3} \label{53}
\end{equation}

Thus, from the above relation one would obtain:%

\begin{equation}
\dot{M}=\frac{4\pi}{3}[\dot{\rho}_\mathrm{0}R_\mathrm{ES}^{3}+3\rho_\mathrm{0}R_\mathrm{ES}^{2}\dot
{R}_\mathrm{ES}]=M[\frac{\dot{\rho}_\mathrm{0}}{\rho_\mathrm{0}%
}+3\frac{\dot{R}_\mathrm{ES}}{R_\mathrm{ES}}] \label{54}
\end{equation}

which simply leads - with the comoving Einstein-Straus radius, i.e. $\dot{R}_\mathrm{ES}/R_\mathrm{ES}=\dot{S}_\mathrm{u}/S_\mathrm{u}=H_\mathrm{0}$ - to the following relation:%

\begin{equation}
\dot{M}/M=[\dot{\rho}_\mathrm{0}/\rho_\mathrm{0}+3H_\mathrm{0}] \label{55}
\end{equation}

With the previous assumption $\dot{M}/M=H_0$ the above relation thus simply requires that:%

\begin{equation}
\dot{\rho}_\mathrm{0}/\rho_\mathrm{0}=-2H_\mathrm{0} \label{56}
\end{equation}

Interestingly enough, as can easily be confirmed, the above relation is
fulfilled for the case that the cosmic mass density $\rho_\mathrm{0}$ scales with
$R_\mathrm{0}^{-2}$. Exactly this relation, however, has been derived as the
unavoidable requirement for a minimum- and constant- energy universe by Fahr
(2004) or Fahr and Heyl (2006a/b). Also Kolb (1989) had found the need for
exactly this relation to be fulfilled for a coasting universe with constant
expansion velocity which also proves to be a constant-energy universe. So,
maybe, PIONEER just gave the first hint for the fact that we are living in a
coasting, economic universe with vanishing total energy and curvature. This even leads to one additional, very interesting connection between vacuum energy decay and mass generation, as we shall show below.

Assuming that there exists a cosmic vacuum energy density $\epsilon_\mathrm{v}%
=\rho_\mathrm{vac}c^{2}$ which is connected with a vacuum pressure $p_\mathrm{vac}$ according to
the following relation (as derived by Fahr and Heyl (2006b)):%

\begin{equation}
p_\mathrm{vac}=-\frac{3-n}{3}\rho_\mathrm{vac}c^{2} \label{57}
\end{equation}

where $n$ is the power index describing the scaling of $\rho_\mathrm{vac}$ with the
scale of the universe, one can then calculate what thermodynamic work is done
by this pressure $p_\mathrm{vac}$, when the Einstein-Straus vacuole with radius
$R_\mathrm{ES}$ expands with the cosmic expansion. Requiring that the energy done by
the vacuum pressure just balances the gain of mass energy of the central
Schwarzschild mass $c^{2}\delta M(t)$ then leads to the following relation:%

\begin{equation}
c^{2}\dot{M}(t)=-(4\pi R_\mathrm{ES}^{2}\dot{R}_\mathrm{ES})p_\mathrm{vac} \label{58}
\end{equation}

which expresses the following context:%

\begin{equation}
c^{2}\dot{M}(t)=3M\frac{\dot{R}_\mathrm{ES}}{R_\mathrm{ES}}\frac{3-n}{3}\frac{\rho_\mathrm{vac}}%
{\rho_\mathrm{mat}}c^{2} \label{59}
\end{equation}

where the central mass has been introduced by $M=(4\pi/3)\rho_\mathrm{mat}R_\mathrm{ES}^{3}$ .
As shown in Fahr and Heyl (2006b) for an economical universe in which the total
energy is constant, it is required that the power index $n$ attains the value
$n=2$. With that one then obtains from the above relation:   %

\begin{equation}
\frac{\dot{M}(t)}{M} = 3\frac{\dot{R}_\mathrm{ES}}{R_\mathrm{ES}}\frac{1}{3}\frac{\rho_\mathrm{vac}}{\rho_\mathrm{mat}} \label{60}
\end{equation}

Reminding that the radius of the Einstein-Straus vacuole expands like: %

\begin{equation}
\frac{\dot{R}_\mathrm{ES}}{R_\mathrm{ES}}=\frac{\dot{S}_\mathrm{u}}{S_\mathrm{u}}=H_\mathrm{0} \label{61}
\end{equation}

one finally arrives at the following most interesting result:%

\begin{equation}
\frac{\dot{M}(t)}{M} = \frac{\dot{S}_\mathrm{u}}{S_\mathrm{u}}\frac{\rho_\mathrm{vac}}{\rho_\mathrm{mat}}=\frac{\rho
_\mathrm{vac}}{\rho_\mathrm{mat}}H_\mathrm{0} \label{62}
\end{equation}

This shows that, for an economical universe with $n=2$, it turns out that the ratio 
$\frac{\rho_\mathrm{vac}}{\rho_\mathrm{mat}}$ is constant and, when the present universe seems
to indicate the value $\frac{\rho_\mathrm{vac}}{\rho_\mathrm{mat}}=(\frac{\rho_\mathrm{vac}}{\rho_\mathrm{mat}%
})_\mathrm{0}\simeq7/3$ (see e.g. Bennet et al., 2003), one astonishingly enough
finds a good explanation of the stellar mass increase required to
explain the PIONEER\ anomaly, namely $\frac{\dot{M}(t)}{M} = \frac{7}{3}H_\mathrm{0}$. However, as shall be shown in section 6.3, it can be revealed as a natural consequence of an economical universe that $\rho_\mathrm{vac}$ should exactly equal $\rho_\mathrm{mat}$, i.e. $\rho_\mathrm{vac} = \rho_\mathrm{mat}$, which would then lead to a perfect agreement with the previous assumption $\frac{\dot{M}}{M} = H_\mathrm{0}$. The main point, however, is that the actual finding for the ratio $\frac{\rho_\mathrm{vac}}{\rho_\mathrm{mat}}$ in an economical universe always leads to the same result at all cosmic times. 

In this context, it may again be highly interesting to see that the appearance of a mass gain
$\dot{M}$ in the ES-vacuole, which can be ascribed to the equivalent of the
work of vacuum pressure at extending the ES-vacuole, can also be ascribed to the change of
gravitational binding energy of the ES-mass in an economic universe where the
total energy vanishes, i.e. $L=E+U=0$, from which follows according to Eqs. (\ref{15}) and (\ref{2101}) for the case of the ES-vacuole:

\begin{equation}
\frac{\mathrm{d}Mc^2}{\mathrm{dt}}= \frac{\mathrm{d}}{\mathrm{dt}}(\frac{4\pi}{3}c^2 R_\mathrm{ES}^{3}\rho_\mathrm{0}) = \frac{\mathrm{d}}{\mathrm{dt}}(\frac{8\pi^{2}G}{15} \rho_\mathrm{0}^{2}R_\mathrm{ES}^{5})=0 \label{100}
\end{equation}

While the time derivative of the expression $\frac{4\pi}{3}c^2 R_\mathrm{ES}^{3}\rho_\mathrm{0}$ leads to Eq. (\ref{55}), the time derivative of the expression $\frac{8\pi^{2}G}{15} \rho_\mathrm{0}^{2}R_\mathrm{ES}^{5}$ results in:

\begin{equation}
\frac{\dot{M}}{M} = \frac{\frac{8\pi^{2}G}{15}\rho_0^{2}R_\mathrm{ES}^{5}(2\frac{\dot{\rho_0}}{\rho_0}+5\frac{\dot{R}_\mathrm{ES}}{R_\mathrm{ES}})}{\frac{8\pi^{2}G}{15}\rho_0^{2}R_\mathrm{ES}^{5}} = (2\frac{\dot{\rho_0}}{\rho_0}+5\frac{\dot{R}_\mathrm{ES}}{R_\mathrm{ES}}) \label{101}
\end{equation}

Since the gain of mass given by Eqs. (\ref{55}) and (\ref{101}), respectively, must be identical, i.e. 

\begin{equation}
\frac{\dot{\rho_0}}{\rho_0}+3\frac{\dot{R}_\mathrm{ES}}{R_\mathrm{ES}} = 2\frac{\dot{\rho_0}}{\rho_0}+5\frac{\dot{R}_\mathrm{ES}}{R_\mathrm{ES}} \label{102}
\end{equation}

one finally derives -  according to Eq. (\ref{61}) - from Eq. (\ref{102}):

\begin{equation}
\frac{\dot{\rho_0}}{\rho_0} = -2 \frac{\dot{R}_\mathrm{ES}}{R_\mathrm{ES}} = -2 \frac{\dot{S_u}}{S_u} = -2 H_0 \label{103}
\end{equation}

This result exactly matches the above formulated requirement in Eq. (\ref{56}).

\section{Implications of the cosmic mass increase for cosmological observations}

In the previous sections we have dicussed several scenarios which could cause an increase of the cosmic mass during the expansion of the universe. Here, a linear increase of the cosmic mass of the universe with increasing scale parameter $S$, or which is identical, a cosmic mass density scaling according to $S^{-2}$ (see section 3.1: economical universe), seems to be a very promising concept. When comparing the present critical mass density of the universe with the theoretically expected value for the mass equivalent of vacuum energy density, this concept appears to be supported by the following remarkable relation, which can hardly be taken as a casual numerical fact (at least from the authors point of view):

\begin{equation}
\frac{\rho_\mathrm{crit}(t_0)}{\bar\rho_\mathrm{vac}} \approx 10^{-122} \approx \frac{{t_\mathrm{Pl}^2 }}{{t_0^2 }} \label{63}
\end{equation}

with $t_\mathrm{Pl}$ the Planck time $\sqrt{G\hbar /c^5 }$, $t_0$ the present age of the universe (13.7 Gyrs), ${\rho_\mathrm{crit}(t_0)}$ the critical density in the present epoch ($\approx 10^{-29} \mathrm{g/cm^3}$), and $\bar\rho_\mathrm{vac}$ the  mass density associated with the theoretically expected vacuum energy density. One assumes the latter to be about half a Planck mass $\frac{1}{2}m_\mathrm{Pl}=\frac{1}{2}\sqrt{\hbar c/G}$ per Planck volume $V_\mathrm{Pl}=\frac{4}{3}\pi r_\mathrm{Pl}^3$, with $r_\mathrm{Pl}=\sqrt{G\hbar/c^3}$.

One might doubt that the above "empirical" relation has a real physical meaning in our observed universe, however, we will show in the following that the above relation and the ratio $\approx 10^{-122}$ are a natural consequence of an universe with a critical density that scales according to $S^{-2}$. Furthermore, this ratio is of special importance with respect to the problem that the "observed" vacuum energy density is smaller exactly by this order of magnitude compared to the vacuum energy density predicted by theory. In this context, the assumption of a cosmic critical density which scales with $S^{-2}$ offers an attractive solution for the presently unsolved problem of this unexplainable high and indigestible vacuum energy density.

\begin{table*}[t] \centering
\begin{tabular}{l|c|c|c|c|c|c|c}
pure case of& $\rho_\mathrm{crit}(S)$ & $S(t)$ & $H(t) = \frac{\dot S(t)}{S(t)}$  & $R_\mathrm{H}(t_0)$ & $M_H(t_0)$ & $R_\mathrm{LH}(t_0)$ & $M_\mathrm{LH}(t_0)$\\
\hline

a.) radiation &$ \propto S^{-4}$ & $
S(t_\mathrm{0})\left( \frac{t}{t_0} \right)^\frac{1}{2} 
$ & $\frac{1}{2t}$ & $2ct_0$& $\frac{c^3}{G}t_0 = \frac{1}{2}\frac{c^2}{G}R_\mathrm{H}$ & $2ct_0$& $\frac{c^3}{G}t_0 = \frac{1}{2}\frac{c^2}{G}R_\mathrm{LH}$\\

b.) matter & $\propto S^{-3}$ & $
S(t_\mathrm{0})\left( \frac{t}{t_0} \right)^\frac{2}{3} $ & $\frac{2}{3t}
$ & $\frac{3}{2}ct_0$ & $\frac{3}{4}\frac{c^3}{G}t_0 = \frac{1}{2}\frac{c^2}{G}R_\mathrm{H}$ & $3ct_0$ & $6\frac{c^3}{G}t_0 = 2\frac{c^2}{G}R_\mathrm{LH}$\\

c.) vacuum & $\rho_\mathrm{vac}=const.$ & $S(t_\mathrm{0}) e^{H(t - t_0 )}$ & const. & const. & $\frac{1}{2}\frac{c^2}{G}R_\mathrm{H} = const.$ & $\propto e^{Ht_0}$ & $\propto e^{3Ht_0} \propto R_\mathrm{LH}^3$\\

d.) economy & $\propto S^{-2}$ & $S(t_\mathrm{0})\frac{t}{t_0}$ & $\frac{1}{t}$ & $ct_0$ & $\frac{1}{2}\frac{c^3}{G}t_0 = \frac{1}{2}\frac{c^2}{G}R_\mathrm{H}$ & $\infty$ & $\infty$\\
\end{tabular}
\caption{\footnotesize{Important cosmic parameters derived from the Friedmann equations are shown as a function of the different scaling of pure cosmic constituents: the density $\rho(S)$, the scale parameter $S(t)$, the Hubble parameter $H(t)$, the radius of the Hubble Sphere $R_\mathrm{H}$ and the mass $M_H$ within the Hubble Sphere, with $t$ the cosmic time, $t_0$ a reference time, e.g. $t_0 = t_\mathrm{today}$, and $S(t_\mathrm{0})$ the associated reference scale parameter.}}
\end{table*}

\subsection{Horizons of the universe}
Before we start the investigation of the physical characteristics of an universe with a cosmic density $\rho \propto S^{-2}$ it is necessary to emphasize the important differences between 1.) the general behaviour of the universe as usually theoretically described with the Friedmann equations applying the time dependent scale parameter $S=S(t)$, and 2.) the observable universe where the finite speed of light $c$ leads to existing boundaries like the \textit{Hubble Sphere} and the \textit{light (or particle) horizon} which have a major impact on the physics of the observable universe. General problems, misunderstandings and misinterpretations related with the different physical horizons in the universe have recently been nicely reviewed by Davis \& Lineweaver (2006).

We begin the examination of the above mentioned boundaries with a look on the first Friedmann equation for an universe with curvature $k=0$:
	
\begin{equation}
\left( {\frac{{\dot S(t)}}{S(t)}} \right)^2  = H(t)^2  = \frac{{8\pi G}}{3}\rho _\mathrm{crit}(t) \label{64}
\end{equation}
	
Here $H$ is again the Hubble parameter, $\dot S$ the time derivative of the scale parameter $S$, and $\rho_\mathrm{crit}$ is the critical density of the universe which can be derived from the above equation as:

\begin{equation}
\rho _\mathrm{crit}(t) = \frac{3H(t)^2}{{8\pi G}} \label{65}
\end{equation}

We now discuss the Hubble sphere which is defined as the distance $R_\mathrm{H}$ where the expansion velocity of the space, i.e. recession velocity of comoving galaxies, equals the velocity of light c:

\begin{equation}
R_\mathrm{H}(t) = \frac{c}{H(t)} \label{66}
\end{equation}

The Hubble Sphere $R_\mathrm{H}$ is usually not assumed to be a real physical border between matter of the universe inside and outside the Hubble Sphere because galaxies with recession velocities $v>c$ outside the Hubble Sphere can later penetrate into the inner region of the Hubble Sphere (where $v<c$) at some time when $H(t)$ has decreased. This is possible because the scale parameter $S$ changes with time according to $t^{1/2}$ or $t^{2/3}$ in cosmological models where the critical density scales according to $S^{-4}$ (pure radiation) or $S^{-3}$ (pure matter), respectively. On the other hand, in these cases the Hubble radius $R_\mathrm{H}$ is with respect to the above equation a linear function of time, i.e. $\propto t$, because the Hubble parameter varies according to $t^{-1}$ for the  cases $S^{-4}$ and $S^{-3}$. Consequently, the Hubble Sphere can "overtake" galaxies at distances which formerly belonged to trans-Hubble-spheric regions to make them part of the inner Hubble Sphere. Table 1 shows the relevant parameters of the universe to clarify this situation for the just discussed cases of a radiation (case a.)) and matter (case b.)) dominated universe, respectively.

In contrast to the Hubble Sphere, the so-called light horizon is definitively a physical rather than a conceptual border since it is the distance that emitted light can have travelled since the beginning of the universe up to the present. Thus, the light horizon is the maximum extension of the visible universe, i.e. the universe an observer is actually physically interacting with. The light horizon $R_\mathrm{LH}$ is given by:

\begin{equation}
R_\mathrm{LH}  = S(t_\mathrm{0})\cdot c\int\limits_0^{t_0 } {\frac{{dt}}{S(t)}} \label{67}
\end{equation}

with $S(t_\mathrm{0})$ being the scale factor at the time $t=t_\mathrm{0}$. For comparison, Table 1 also shows the explicite expressions for the light horizon $R_\mathrm{LH}$ and the associated mass contents $M_\mathrm{LH}$, respectively. In the following we shall discuss the physical meaning of this table in detail.

\subsection{The standard cosmological model}

According to the standard model of cosmology the universe started with a radiation dominated era ($\rho_\mathrm{crit} \propto S^{-4}$) lasting for approx. 380 kyears, followed by a matter dominated epoch ($\rho_\mathrm{crit} \propto S^{-3}$). Nowadays, the universe is assumed to be dominated by the prevailing vacuum energy which is considered to be a  constant ($\rho_\mathrm{crit} = const.$). These 3 epochs are represented by the cases a.) to c.) in Table 1. It is remarkable that in all cases the resulting mass content $M_\mathrm{H}$ of the Hubble Sphere follows the same physical law, i.e. yielding a linear increase of the mass $M_\mathrm{H}$ of the universe with the extension $R_\mathrm{H}=c/H$:

\begin{equation}
M_H  = \rho _\mathrm{crit} \frac{4\pi}{3}R_\mathrm{H}^3  = \frac{3H^2}{8\pi G}\frac{4\pi }{3}\frac{c^3}{H^3} =  \frac{1}{2}\frac{c^2}{G}R_\mathrm{H} \label{68}
\end{equation}

The reason for this behaviour is, however, the already mentioned fact, that the Hubble Sphere is no physical border for the cases a.) and b.), since matter from beyond the Hubble Sphere can intrude into the inner Hubble region after some time. In the vacuum dominated epoch (case c.)) the Hubble parameter $H$ has a constant value due to the assumed constant energy density of the vacuum. Therefore, $R_\mathrm{H}$ is constant and thus $M_\mathrm{H} = const$, too. The situation is more or less similar with respect to the mass $M_\mathrm{LH}$ within the light horizon (Table 1), which is given by: 

\begin{equation}
M_\mathrm{LH}  = \rho _\mathrm{crit} \frac{4\pi}{3}R_\mathrm{LH}^3 \label{69}
\end{equation}

Now, in all 3 cases the mass within the light horizon increases with $R_\mathrm{LH}$ since more and more space becomes visible, the mass content of which (photons, matter, vacuum energy) contributes to the observable universe.

Much more interesting is the recognition that in the standard model of cosmology the observable universe started its  radiation dominated era with half a Planck mass, i.e. $\frac{1}{2}m_\mathrm{Pl}$. This can also be derived from Table 1 which reveals that during the radiation dominated area the Hubble Sphere $R_\mathrm{H}$ and the light horizon were identical ($R_\mathrm{H} = R_\mathrm{LH}$). Thus, the mass $M_{LH}$ of the observable universe at the very beginning, i.e. the Planck length $r_\mathrm{Pl}= \sqrt {G\hbar/c^3} = R_\mathrm{H} = R_\mathrm{LH}$, can be calculated as:

\begin{equation}
M_\mathrm{LH}(r_\mathrm{Pl} ) = \frac{{c^2 }}{{2G}}r_\mathrm{Pl}  = \frac{1}{2}\sqrt {\frac{{\hbar c}}{G}}  = \frac{1}{2}m_\mathrm{Pl} \label{70}
\end{equation}

This means, according to the cosmological principle, that any arbitrary observer in the "standard model" universe started with a mass content $\frac{1}{2}m_\mathrm{Pl}$ of his point-related universe and that the mass he was interacting with during the following expansion of the universe was steadily increasing with time. 

For completeness we also mention the critical densities $\rho_\mathrm{crit, rad}$ and $\rho_\mathrm{crit, mat}$ during the radiation and the following matter dominated area:

\begin{equation}
\rho _\mathrm{crit,rad}(t)  = \frac{{3H_\mathrm{rad}^2(t) }}{{8\pi G}} = \frac{3}{{8\pi G}}\frac{1}{{4t^2 }} \label{71}
\end{equation}

\begin{equation}
\rho _\mathrm{crit,mat}(t)  = \frac{{3H_\mathrm{mat}^2(t) }}{{8\pi G}} = \frac{3}{{8\pi G}}\frac{4}{{9t^2 }} \label{72}
\end{equation}

with $H_\mathrm{rad}=\frac{1}{2t}$ and $H_\mathrm{mat}=\frac{2}{3t}$ as the associated Hubble parameters during the respective epochs (see Table 1).

\subsection{The economical universe}

The basic idea of an so-called "economical" or "zero-energy" universe with a the critical density which scales according to $S^{-2}$ (cases d.) in Table 1) has been introduced in section 3.1. We now want to discuss the most important consequences of such an economical universe, in particular with a focus on the problem of the vacuum energy density, the theoretical value of which is about $10^{122}$ times higher than the value it should have to be conciliant with present observational results. This discrepancy could be easily eliminated with the assumption of a cosmic mass (or energy) density scaling with $S^{-2}$ and the associated solution to the presently unsolved $10^{122}$-problem shall be quantified in more detail in the following.

A look on Table 1 reveals the interesting fact that for $\rho_\mathrm{crit} \propto S^{-2}$ the mass within the Hubble Sphere follows the same law as given in Eq. (\ref{68}), i.e. a linear scaling with $R_\mathrm{H}$. This linear mass increase, however, cannot be related to matter outside the Hubble Sphere which crosses the Hubble border at some time to intrude into the inner Hubble region because the linear increase of $R_\mathrm{H}$ on one hand, and the linear increase of $S(t)$ with time on the other hand, does not allow for a crossing of trans-Hubble matter through the Hubble Sphere. In other words, for case d.) the Hubble Sphere is a real physical border for matter, and thus matter from outside the Hubble Sphere can never reach the inner Hubble Sphere to cause a mass increase there. Therefore, the mass growth must be due to the creation of new mass within the Hubble Sphere, such that the mass is again given by the expression (see also Table 1 and Eq. (\ref{68})):

\begin{equation}
M_H  = \frac{1}{2}\frac{c^2}{G}R_\mathrm{H} \label{73}
\end{equation}

or, with $R_\mathrm{H} = ct$ according to Eq. (66):

\begin{equation}
M_H  = \frac{1}{2}\frac{{c^3 }}{G}t \label{74}
\end{equation}

The creation of the mass could be easily ascribed to quantum mechanical effects. A look at the uncertainty principle ${\hbar/2} \approx \Delta E\Delta t$ offers the possibility of the virtual apperarance of half a Planck mass within a time interval $\Delta t = t_\mathrm{Pl}$, i.e. Planck time:

\begin{equation}
{\frac {\hbar} {2}} \approx \Delta E\Delta t = \Delta mc^\mathrm{2} t_\mathrm{Pl}  = \Delta mc^\mathrm{2} \sqrt {{\frac {G\hbar} {c^\mathrm{5}}}}, \label{75}
\end{equation}

and thus:

\begin{equation}
\Delta m = {\frac {1} {2}}\sqrt {{\frac {\hbar c} {G}}}  = {\frac {1} {2}}m_\mathrm{Pl}. \label{76}%
\end{equation}

This virtual "half Planck mass" may be lifted up to the real world of the universe, if its rest mass energy is compensated to zero by additional gravitational binding energy. The negative gravitational binding energy which each additional mass is subject to in the expanding universe may lead to a vanishing change of the total energy of the whole universe (see again section 3.1: economical universe). Thus, the mass increase of the universe could have its reason in virtual Planck masses which become real and which contribute over the lifetime of the universe to the total mass $M_H$ with a "production rate" of half a Planck mass per each time interval $t_\mathrm{Pl}$:

\begin{equation}
M_\mathrm{H}  = {{\frac {1} {2}}{\frac {m_\mathrm{Pl}} {t_\mathrm{Pl}}}}t = {\frac {1} {2}}{\frac {c^\mathrm{3}} {G}}t \label{77}
\end{equation}

or, which is identical, with a production rate of half a Planck mass per size increment by one "Planck length $r_\mathrm{Pl}$" at the expansion of the universe:

\begin{equation}
M_\mathrm{H}  = {{\frac {1} {2}}{\frac {m_\mathrm{Pl}} {ct_\mathrm{Pl}}}}ct = {{\frac {1} {2}}{\frac {m_\mathrm{Pl}} {r_\mathrm{Pl}}}}ct = {\frac{1}{2}\frac {c^\mathrm{2}} {G}}R_\mathrm{H}. \label{78}%
\end{equation}

where we have again used the relation $R_\mathrm{H} = ct$ which applies to a universe with $\rho_\mathrm{crit} \propto S^{-2}$. According to this equation, the initial cosmic mass $M_\mathrm{H}(t_\mathrm{Pl})$ at the beginning of the universe is $\frac{1}{2}m_\mathrm{Pl}$, which is valid also for the cases a.) to c.), see Eq. (\ref{70}).

The existence of a constant vacuum energy density in the standard model of cosmology is based on a constant pressure $p_\mathrm{vac}$ which is associated with the expansion of the universe and which is given by (see Peebles and Ratra 2003):

\begin{equation}
p_\mathrm{vac}=-\rho_\mathrm{vac}c^2 \label{79}
\end{equation}

This vacuum energy density can be interpreted as the store of an enormous amount of virtual energy, i.e. the equivalent of the rest energy of half a Planck mass per Planck volume. The arising problem of the standard model universe now is that the gain of new space during the expansion results in the gain of exactly this rest energy of half a Planck mass each gained Planck volume, which finally leads to the afore mentioned well-known $10^{122}$-discrepancy.

This problem does not come up if $\rho_\mathrm{crit} \propto S^{-2}$ is applied, because the economical universe is controlled by the conservation of energy. Thus, the release of the rest energy of half a Planck mass during the expansion of an economical universe can take place, if and only if the energy condition allows for such a release, i.e. if the total energy remains zero due to the compensation of the released energy by the negative gravitational binding energy. This condition is fulfilled for an expansion rate of one Planck length each time interval with a duration of one Planck time which leads to the Eqs. (\ref{77}) and (\ref{78}). In this sense, the quantum mechanical production of real Planck masses in the economical universe is strongly coupled to and restricted by an adequate expansion dynamics which guarantees a vanishing total energy at the lapse of cosmic time. As just argued, this is realized for a critical density $\rho_\mathrm{crit} \propto S^{-2}$ which is according to Eq. (\ref{65}) given by:

\begin{equation}
\rho _\mathrm{crit}(t)  = \frac{{3H^2(t) }}{{8\pi G}} = \frac{3}{{8\pi G}}\frac{1}{{t^2 }} \label{80}
\end{equation}

where we have used $H(t)=\frac{1}{t}$ for $\rho \propto S^{-2}$ as shown in Table 1. In the economical universe, this critical density is nothing else but materialized vacuum energy the mass density of which amounts at the beginning of the universe, i.e. $t=t_\mathrm{Pl}$, to:

\begin{equation}
\rho _{crit} (t_{Pl} ) = \bar\rho _{vac}  = \frac{3}{{8\pi G}}\frac{1}{{t_{Pl}^2 }} = \frac{{\frac{1}{2}m_{Pl} }}{{\frac{4}{3}\pi r_{Pl}^3 }} \label{81}
\end{equation}

where $\bar\rho_\mathrm{vac}$ is again the theoretical value for the equivalent mass density of the vacuum energy. The above equation expresses that an economical universe does not consist of a magic "vacuum energy component" which co-exists side by side with the matter component as is the case in the standard model universe, but the vaccum energy of the economical universe manifests itself as matter which is released by the vacuum, thereby respecting the conservation of energy. This is the reason why an economical universe does not face the $10^{122}$-discrepancy. But this is only one advantage of an universe with $\rho_\mathrm{crit} \propto S^{-2}$. Other important features will be listed in the next section.

\section{Important characteristics of an economical universe}

As just shown above both, the economical universe and the radiation dominated universe in the standard model start with half a Planck mass at the beginning of their expansion. According to Eq. (\ref{74}), today ($t_\mathrm{0}=\mathrm{13.7 Gyrs}$) the mass of the economical universe amounts to:

\begin{equation}
M_\mathrm{H}  = {\frac {1} {2}}{\frac {c^\mathrm{3}} {G}}t_0 \approx 10^{53} \mathrm{kg} \approx10^{80}\mathrm{m}_\mathrm{prot} \label{82}
\end{equation}

The above values are consistent with standard estimations, e.g. what concerns the visible universe to consist of about $10^\mathrm{11}$ galaxies with $10^\mathrm{11}$ solar-type stars each. We emphasize that we are talking about the mass within the Hubble Sphere, i.e. the amount of mass which grows due to the described effect of mass release by the vacuum, since no mass from outside the Hubble border can intrude the inner Hubble Sphere.

According to Eq. (\ref{80}) the associated critical density today, i.e. at $t=t_0 = 13.7$ Gyrs, simply yields:

\begin{equation}
\rho_\mathrm{crit}(t_0) = \frac{3}{{8\pi G t^2_\mathrm{0}}} \approx \mathrm{10^{-26} \frac{kg}{m^3} \approx 10^{-29} \frac{g}{cm^3}} \label{83}
\end{equation}

This is in very good agreement with the presently assumed value of the critical density. Furthermore, the relation between the age $t_0$ of our universe and the Hubble parameter is simply given by (see Table 1):

\begin{equation}
t_\mathrm{0}= \frac{1}{H(t_0)} \Leftrightarrow H(t_0)= \frac{1}{t_\mathrm{0}} \label{84}
\end{equation}

Again, it is remarkable that the presently accepted age of the universe and the generally accepted value of the Hubble parameter ($\mathrm{72 km/s/Mpc}$) are in perfect agreement since $H(t_0)t_0 \approx 1$ for $\rho_\mathrm{crit} \propto S^{-2}$.

We now come back to Eq. (\ref{63}) and the speculation that this relation is not an artefact but has a physical background. This can now be proved with the Eqs. (\ref{80}) and (\ref{81}) when calculating the ratio of the present critical density and the critical density at the Planck time, respectively, which leads to:

\begin{equation}
\frac{{\rho _\mathrm{crit} (t_0)}}{{\rho _\mathrm{crit}(t_\mathrm{Pl})}}= \frac{{\rho _\mathrm{crit} (t_0)}}{{\bar\rho _\mathrm{vac}}} \mathop  = \frac{{t_\mathrm{Pl}^2 }}{{t_0^2 }} 
\mathop  \approx 10^{-122} \label{85}  
\end{equation}

These results are the reason why the authors believe that the vacuum energy density is not a constant but scales according to $S^{-2}$ or, i.e. with $S \propto t$, scales according to $t^{-2}$, respectively. Such a scaling has not only the advantage that the $10^{122}$-problem of the vacuum energy can be dissolved, but also that the universe - even without inflation - does not face a horizon problem since the light horizon according to Eq. (\ref{67}) reaches infinity for $\rho \propto S^{-2}$ with a universe starting its evolution at a time $t \approx 0$.

Finally, there is no reason to be concerned by the linear time dependence $S \propto t$ of the scale factor causing dramatic changes in the timeline of the nucleosynthetic processes in the early universe. At first glance, one would expect significant impacts on the nucleosysnthesis since the cosmic energy density scales with $S^{-4}$ during the radiation dominated epoch of the standard model, while the economical universe scales according to $S^{-2}$. However, if one looks at the densities in the Eqs. (\ref{71}) and (\ref{80}) one recognizes that the cosmic densities follow the same law with respect to the cosmic time, i.e. $\rho_\mathrm{crit} \propto \frac{1}{4t^2}$ and $\frac{1}{t^2}$ for the cases $S^{-4}$ and $S^{-2}$, respectively. Thus, the critical densities of the radiation dominated universe in the standard model and the economical universe differ by a factor 4 only. In this context we recall the usual "time-temperature relationship" for a mixture of interacting relativistic bosons and fermions in the early universe (G\"onner 1994), i.e. the expression for the temperature $T$ of this mixture at a given cosmic time $t$:

\begin{equation}
t \propto \frac{1}{T^2 } \label{86}
\end{equation}

This equation follows from the relations between the Hubble parameter, the total energy density $\epsilon_\mathrm{tot}$ of the mixture of the concerned relativistic particles, and the temperature of this mixture, respectively:

\begin{equation}
H^2(t) \propto \epsilon_\mathrm{tot} \propto T^4  \label{87}
\end{equation}

We see, that with $H(t)=\frac{1}{2t}$ for $S^{-4}$ and $H(t)=\frac{1}{t}$ for $S^{-2}$ the ratio of the temperatures at a cosmic time $t$ in the early universe is given by:

\begin{equation}
\frac{T(S^{-2})}{T(S^{-4})}= \frac{\frac{1}{\sqrt{t}}}{\frac{1}{\sqrt{2t}}}= \sqrt{2} \label{88}
\end{equation}

This small difference of a factor $\sqrt{2}$ will have only very little impacts on the timeline of the classical nucleosynthesis, if any, as suggested by an investigation of the impacts of a concordant "freely coasting" universe (Kolb 1989).

\section{Conclusions}
As we believe, we have shown in the foregoing sections of this paper that the
concept of a constant cosmic vacuum energy density, though usually applied in
modern cosmology of these days, is unsatisfactory for many reasons,
especially, however, for one very basic reason: This concept namely would
claim for a physical reality that acts upon spacetime and matter dynamics
without itself being acted upon by spacetime or matter. This fact is
demonstrated from many different aspects and then analysed with respect to its
numerous cosmological consequences. We first have looked into the cosmological
literature of the more distant past where cosmic mass generation mechanisms
had been formulated in order to describe a steady state universe (see Hoyle,
1948 or Bondi and Gold, 1948) or scale-covariant universes (see Hoyle and
Narlikar, 1966a/b) and can show from our analysis that various, theoretically
described forms of mass generation in the universe, to some surprise, lead to
terms in Einstein`s general relativistic field equations which in many
respects are analogous to, or can even be replaced by terms arising from
vacuum energy. 

We then analyse other cosmological attempts in the literature of the more
recent past  where it could be demonstrated that gravitational cosmic binding
energy acts as negative cosmic mass energy density and, as a surprise, again
can be shown to call for additional terms similar to those written for cosmic
vacuum energy density. As it clearly turns out from our studies here the
cosmic actions of vacuum energy, gravitational binding energy and mass
creation are obviously closely related to eachother. 

Based on these results we feel encouraged to suggest that the action of vacuum
energy on cosmic spacetime necessarily and simultaneously leads to a decay of
vacuum energy density, a decrease of cosmic binding energy, and a creation of
mass in the expanding universe. We can demonstrate that only under these
auspices an expanding universe with constant total and minimal energy, the
so-called economic universe, can exist. In such a universe both cosmic mass
density and cosmic vacuum energy density are decreasing according to
$(1/S^{2})$, where $S$ denotes the characteristic scale of the universe. Only
under these conditions the origin of the present massive universe from an
initially empty one, i.e. a complete vacuum, reveals possible. The incredibly
huge vacuum energy density which quantumfield theoreticians do derive is the
guiding number, but it meanwhile has decayed during the expansion of the
universe to just that small value (n.b. a factor of $10^{-122}$)  which is
conciliant with the behaviour of the present universe, but reappears in the
energy density of created cosmic matter.

In conclusion of all these above considerations one nevertheless can frankly
say that up to now there is surely a lack of a rigorous formulation for the
transition of vacuum fluctuations into real masses, nevertheless there is at
least the idea first discussed by Hawking (1975) that treating the
quantummechanics of particles and antiparticles in the neighborhood of
blackholes reveals the appearance of real particles, i.e. the decay of a pure
vacuum into a mass-loaded vacuum around these blackholes. The so-called
Hawking radiation in this respect is nothing else but a materialisation of
vacuum energy occuring in strong gravitational fields. Perhaps in this respect
the expanding universe also represents a form of a time-dependent
gravitational field with strong gradients both in time and space which induces matter creation through the embedded
time-dependent quantummechanical wavefunctions of particles.

\end{document}